# Two-Phase Change in Thickness and Viscoelasticity of Polyelectrolyte Multilayers Swollen with Ionic Liquid Solution

*Nagma Parveen[1,2,†,*], Pritam Kumar Jana[3] and Monika Schönhoff[1]*

[1]Institute of Physical Chemistry, University of Muenster, Corrensstr. 28/30, 48149 Münster, Germany

[2]NRW Graduate School of Chemistry, University of Muenster, Wilhelm-Klemm-Str. 10, D-48149 Münster, Germany

[3]Université Libre de Bruxelles, Interdisciplinary Center for Nonlinear Phenomena and Complex Systems, Boulevard du Triomphe, 1050 Brussels, Belgium

[†]*current address:* Laboratory for Photochemistry and Spectroscopy, Molecular Imaging and Photonics Division, Department of Chemistry, KU Leuven, B-3001 Leuven, Belgium

**Abstract**

Polyelectrolyte multilayers (PEM) in combination with good electrolyte solvents, e.g., ionic liquids (ILs) are potential candidates for the new generation of electrochemical separators. Swelling PEM with aqueous IL solutions is one way to incorporate IL in PEM. In addition to quantifying the IL uptake in PEM, physical characterization of the swollen PEM is essential for their potential applications and to uncover the underlying molecular process of the swelling. In this article, thickness and viscoelasticity of PEI(PSS/PADMAC)$_4$PSS and PEI(PSS/PAH)$_4$PSS multilayers were determined upon uptake of aqueous solutions of 1-Ethyl-3methylimidazolium chloride, a water-soluble IL. This was done by employing the Kelvin-Voigt model in combination with the frequency and dissipation shifts measured using quartz crystal microbalance with dissipation technique. The modeled thickness and viscoelasticity of the multilayers change in two phases with increasing IL concentration. In the first phase, a small increase in PEM thickness accompanies a two to three-fold increase in PEM viscoelasticity, whereas the changes are other ways around in the second phase. Two phenomenon contributes to the two-phase changes: 1) the exchange between hydrated ions of IL and in-situ water of PEM influencing the overall ion and water content in PEM and 2) the hydrophobic and electrostatic interactions between incorporated IL and PEM influencing the structural integrity of PEM.

**Introduction**

Polyelectrolyte multilayers (PEM) are three-dimensional films and formed upon adsorption of oppositely charged polyelectrolytes in a layer-by-layer (LbL) fashion[1]. Thickness and elasticity of PEM can be controlled within a range of nm to cm and 5 MPa to 1 GPa, respectively. This is possible by simply varying the number of deposited layers[2,3] and/or the molecular properties of composite polyelectrolytes, e.g., charge density, persistence length, molecular structure [4, 5, 6]. The mechanically robust yet thin feature of PEM makes them an attractive candidate as coating materials[7], substrates for biological molecules/systems[8], and more recently, as electrochemical separators[9]. For the latter application, PEM must be conductive of ions and that can be achieved by combining PEM with an electrolyte solvent, e.g., ionic liquid (IL). One way to combine PEM with IL is to incorporate ions of IL in an already assembled multilayer. Earlier we have shown that PEM can be swelled with IL solution and the uptake of IL in PEM can be controlled by varying IL's hydrophobicity[10] and the charge excess in PEM[11]. The IL incorporation and the resulting swelling of PEM are likely to change the thickness and mechanical strength, e.g., viscosity, elasticity of PEM. These physical properties of the swollen PEM describe whether the films are thin yet robust enough for the potential applications. Moreover, a thorough analysis of the changes in the thickness and viscosity/elasticity may uncover the underlying molecular process of the PEM swelling with IL solutions. This molecular analysis will elevate the fundamental understanding of the swelling of PEM and its consequences in general.

In this work, we implemented the Kelvin-Voigt model to determine thickness and viscoelasticity of PEM swelled with IL solutions. The modeling was done by exploiting the piezoelectric response of quartz crystal microbalance with dissipation (QCM-D) technique. The acoustic wave propagated through a quartz surface in QCM-D is sensitive to the thickness and mechanical properties, e.g., elasticity of a film adsorbed on the quartz surface. This means that thickness and mechanical properties of the adsorbed film can be extracted from the frequency ($\Delta f$) and dissipation ($\Delta D$) shift of the acoustic wave in QCM-D[12]. Unlike thickness, the mechanical property of a viscoelastic film requires a model which can suitably describe the viscoelasticity of the film. One popular model is the Kelvin-Voigt model which is applicable for relatively rigid viscoelastic films[12,13]. Because of the non-covalent cross-linking of the oppositely charged polyelectrolytes in PEM, the structural rigidity of PEM is relatively high[14-15]. Hence, the Kelvin-Voigt model is suitable to describe the viscoelasticity of PEM. In this work, we employed the Kelvin-Voigt model to deconvolute thickness, viscosity and elasticity of PEM from its QCM-D shifts. In particular, we have performed the modeling on two charge-neutral PEM, i.e., PEI(PSS/PDADMAC)$_4$PSS and PEI(PSS/PAH)$_4$PSS upon swelling with aqueous solutions of 1-Ethyl-3methylimidazolium chloride which is a water-soluble IL. The QCM-D data used for the modeling are published in our earlier work where the reversible swelling phenomenon of PEM with aqueous solutions of the IL was eastbalished[10]. In this work, we extracted the thickness and viscoelasticity of the swollen multilayers to explain the mechanism behind the observed swelling trend. We found out that the thickness and viscoelasticity of PEM change in two phases upon swelling the multilayers with increased IL concentration. In the first phase, the thickness of PEM increases weakly accompanied by a two to three-fold increase of the viscoelasticity of PEM. In the second phase, PEM thickness increases sharply accompanied by a minimal change of the PEM viscoelasticity. This contrasting trend in the thickness and viscoelasticity of the swollen PEM indicates that the swelling is not a direct consequence of increasing IL uptake in PEM, instead, it is driven by an exchange between IL solution and in-situ water of PEM. The extent of this exchange and the molecular interactions between the incorporated ions of IL and PEM control the observed changes in the thickness and mechanical properties of PEM.

**Method**

**Kelvin-Voigt Modeling.** The $\Delta f$ shift of QCM-D is typically analyzed to express or determine the surface coverage of a film adsorbed on a quartz surface. In case of a solid/rigid film, the determined surface coverage can be converted to the film thickness from a known density of the film. However, PEM are solid-like viscoelastic films that is why such surface coverage to thickness conversion is not straightforward for PEM. Moreover, the liquid medium, e.g., water or IL solution in and around PEM contribute to the $\Delta f$ shift in QCM-D. Energy dissipation of the acoustic wave generated in QCM-D, known as simply dissipation ($\Delta D$), is another parameter of QCM-D. It expresses the liquid-like environment, or in other words the softness of the adsorbed film. This means that both $\Delta f$ and $\Delta D$ shift of PEM are dependent on the viscoelasticity of PEM as well as the viscosity of the immersion liquid. We chose the Kelvin-Voigt model to express the viscous and elastic component of the viscoelastic PEM. Equations 1 and 2 present the exact dependence of the $\Delta f$ and $\Delta D$ shift on different physical parameters of the adsorbed film and immersion liquid in QCM-D[12].

$$\Delta f \approx -\frac{1}{2\pi \rho_0 d_0} \left\{ \frac{\eta_2}{\delta_2} + \left[ d_1 \rho_1 \omega - 2d_1 \left(\frac{\eta_2}{\delta_2}\right)^2 \frac{\eta_1 \omega^2}{\mu_1^2 + \omega^2 \eta_1^2} \right] \right\} \qquad (1)$$

$$\Delta D \approx \frac{1}{2\pi f \rho_0 d_0} \left\{ \frac{\eta_2}{\delta_2} + \left[ 2d_1 \left(\frac{\eta_2}{\delta_2}\right)^2 \frac{\mu_1 \omega^2}{\mu_1^2 + \omega^2 \eta_1^2} \right] \right\} \qquad (2)$$

where $f$, $\rho$, $\mu$, $\eta$ and $d$ designate resonant frequency (for the given overtone), density, elastic modulus, viscosity and thickness, respectively and $\delta = \sqrt{\frac{2\eta}{\rho\omega}}$ ($\omega$ is a complex number). The subscripts 0, 1, 2 correspond to the quartz crystal, adsorbed film and immersion medium (water or IL solution), respectively.

In practice, the theoretical $\Delta f$ and $\Delta D$ shifts of PEM were computed using Equations 1 and 2 which account the viscoelasticity of PEM according to the Kelvin-Voigt model[12]. The equations take into account parameters of both PEM and the immersion liquid, i.e., either water of IL solution. To compute the equations, we kept some of the parameters as fixed and others as free parameters. The fixed parameters were the density and viscosity of the immersion liquid and the density of PEM. The free parameters were the thickness, viscosity and elastic modulus of PEM. Parameters of the quartz crystal are constant and were taken into account while performing the modeling in Q-Tools software (version 3.1.25.604, Biolin Scientific, Sweden). Table 1 lists the values and range of the fixed and free parameters for the modeling, respectively. The computed QCM-D shifts were then compared with the measured QCM-D shifts at the respective IL concentration. A least-square fitting method was used to find the best fit of the measured and modeled QCM-D shifts, and the value of the free parameters at the best-fit was the output/result of the modeling. The computation and fitting were performed over five different overtones of QCM-D frequency, ensuring a robust modeling (Figure S2 and S3 in the Supporting Information).

|  | Fixed parameters | | | Free parameters | | |
|---|---|---|---|---|---|---|
| Immersion liquid | Liquid Density, $\rho_2$ (kg m$^{-3}$) | Liquid Viscosity, $\eta_2$ (kg m$^{-1}$ s$^{-1}$) | PEM Density, $\rho_1$ (kg m$^{-3}$) | PEM Thickness, $d_1$ (nm) | PEM Viscosity, $\eta_1$ (kg m$^{-1}$ s$^{-1}$) | PEM Elastic modulus, $\mu_1$ (MPa) |
| Water | 1000 | 0.001 | 1070 | 0.1 to 1000 | 0.005 to 1 | 1 to 1000 |
| IL solution | 1000 | In Figure 1 | 1070 | 0.1 to 1000 | 0.005 to 1 | 1 to 1000 |

**Table 1.** Fixed and free parameters employed for the Kelvin-Voigt modeling of PEI(PSS/PADMAC)$_4$PSS and PEI(PSS/PAH)$_4$PSS multilayers with either water or IL solution as the immersion liquid. As tabulated, a well-reported mass-density (1070 kg m$^{-3}$) of PEM assembled from synthetic polyelectrolytes[16,17] was chosen for the modeling. Values of other fixed parameters are either standard or measured.

**Results**

**Kelvin-Voigt modeling of hydrated PEM.** PEM, adsorbed on the quartz surface of QCM-D, are in hydrated state and the last (upper most) polyelectrolyte layer of PEM is in direct contact with the bulk water. The Kelvin-Voigt modeling of the hydrated PEM was performed taking into account the fixed parameters provided in Table 1. The thickness, viscosity and elastic modulus of PEI(PSS/PDADMAC)$_4$PSS and PEI(PSS/PAH)$_4$PSS multilayers (see Table 2) determined from the modeling are in agreement with the reported thickness and elasticity of similar PEM determined using other experimental techniques/methods[18-21]. Our results presented in Table 1 show that the PSS/PDADMAC film is thicker and more viscoelastic in comparison to PSS/PAH, indicating a higher stress resistance of the former film.

| PEM | Thickness, $d_1$ (nm) | Viscosity, $\eta_1$ (kg m$^{-1}$ s$^{-1}$) | Elastic modulus, $\mu_1$ (MPa) |
|---|---|---|---|
| PSS/PDADMAC | 36.78 ± 0.88 | 0.058 ± 0.012 | 4.79 ± 1.84 |
| PSS/PAH | 25.33 ± 0.55 | 0.023 ± 0.006 | 1.02 ± 0.27 |

**Table 2.** Thickness, viscosity and elastic modulus of hydrated PEI(PSS/PADMAC)$_4$PSS and PEI(PSS/PAH)$_4$PSS multilayers obtained from the Kelvin-Voigt modeling of the multilayers.

**Kelvin-Voigt modeling of PEM swollen with ionic liquid solution.** In QCM-D, $\Delta f$ and $\Delta D$ shift of PEM decreases and increases, respectively upon injection of an IL solution to the PEM (Figure S1 in the Supporting Information). Even after the bulk-effect correction, the shifts in $\Delta f$ and $\Delta D$ are substantially high which means a mass uptake in PEM upon addition of IL solution[10]. More precisely, the uptake of hydrated ions of the IL in PEM results in a swelling of PEM. The Kelvin-Voigt modeling of the swollen multilayers was performed using the given parameters of Table 1. While modeling we observed that the least-square fit of the computed QCM-D shifts was most sensitive to the viscosity of IL solutions ($\eta_2$), see Figure S2 of the Supporting Information. This is why unlike the other two fixed parameters, i.e., density of PEM and IL solution (Table 1), precise viscosity of IL solution at a given concentration was used in our modeling. For this, the viscosity of IL solutions measured using Rheometer (Figure 1) was chosen as the fixed viscosity parameter ($\eta_2$) for the modeling of the swollen PEM. In most cases, a best-fit modeling was achieved using a $\eta_2$ value which shifts only 3 to 5 % from the measured viscosity of the respective IL concentration (Figure 1).

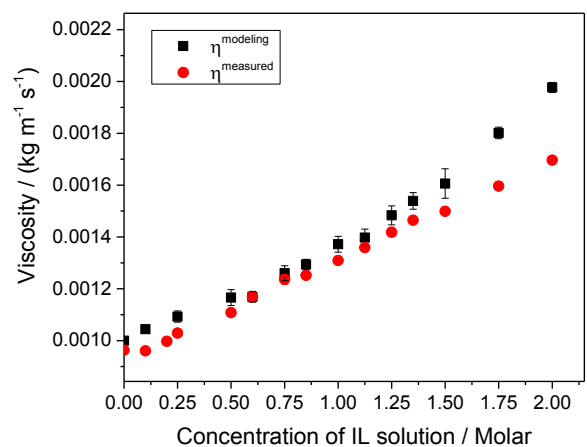

**Figure 1.** The viscosity of IL solutions at the best-fit modeling ($\eta^{modeling}$) of swollen PEM and measured ($\eta^{measured}$) using Rheometer in the rotation method at given IL concentrations. The $\eta^{modeling}$ is $\eta_2$ parameter while modeling the swollen PEM.

**Thickness of PEM swollen with ionic liquid solution.** The thickness values obtained from the modeling of the swollen PEM shows a small followed by a steep rise with increasing IL concentration, see Figure 2. This two-phase increase eventually follows a thickness drop in case of PSS/PDADMAC multilayers, indicating a partial dissolution of the film (Figure 2a). Such dissolution of PSS/PAH multilayers was not obvious. However, the large error bars in the thickness of PSS/PAH at ≥2.5 M IL concentration are an indication of the commencement of the film dissolution. For both PSS/PDADMAC and PSS/PAH multilayers, the thickness of the swollen film reaches up to ~1.5 times of the hydrated film thickness. This means that with respect to the thickness change the charge-neutral PEM assembled from synthetic polyelectrolytes exhibit a similar response upon IL incorporation.

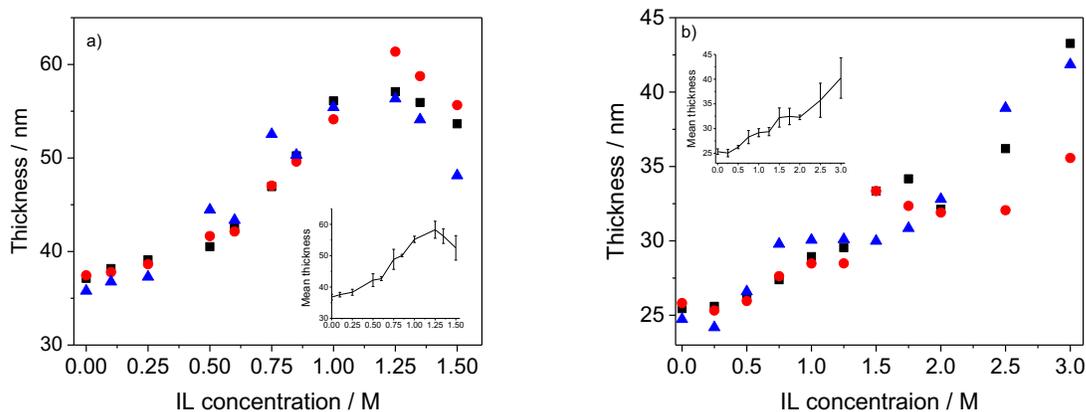

**Figure 2.** Modeled thickness of a) PEI(PSS/PDADMAC)$_4$PSS and b) PEI(PSS/PAH)$_4$PSS multilayers upon swelling with IL solutions of different molar concentrations. Insets show the average thickness values of the given three datasets and the corresponding standard deviations. The lines are drawn to guide the eyes.

**Viscoelasticity of PEM swollen with ionic liquid solution.** The viscosity and elastic modulus of the swollen multilayers also show a two-phase change with increasing IL concentration (Figure 3). In the two-phases, the viscoelasticity increases two to four-fold followed by a decrease to the initial or slightly lower than the initial viscoelasticity of the hydrated PEM. Interestingly, the changes in viscoelasticity and thickness of PEM swollen with increasing IL concentration are just the opposite of each other (Figure 2 and 3). This behavior is easy to visualize in Figure 4 where the thickness and viscosity of swollen PSS/PADMAC are plotted for one dataset.

Together, our modeling data show that PEM thickness changes minimally with a two to four-times increase in their viscosity and elastic modulus in the first-phase of the swelling. Whereas the multilayers stretch up to ~1.5 times of the hydrated film thickness with a minimal change in their viscosity and elastic modulus in the second phase of the swelling. Although the error bars of the average viscoelasticity data are relatively large (inset of Figure 3) compared to that of the thickness data, still the above-mentioned trends in the thickness and viscoelasticity of the individual dataset are well-traceable (see in Figure 4).

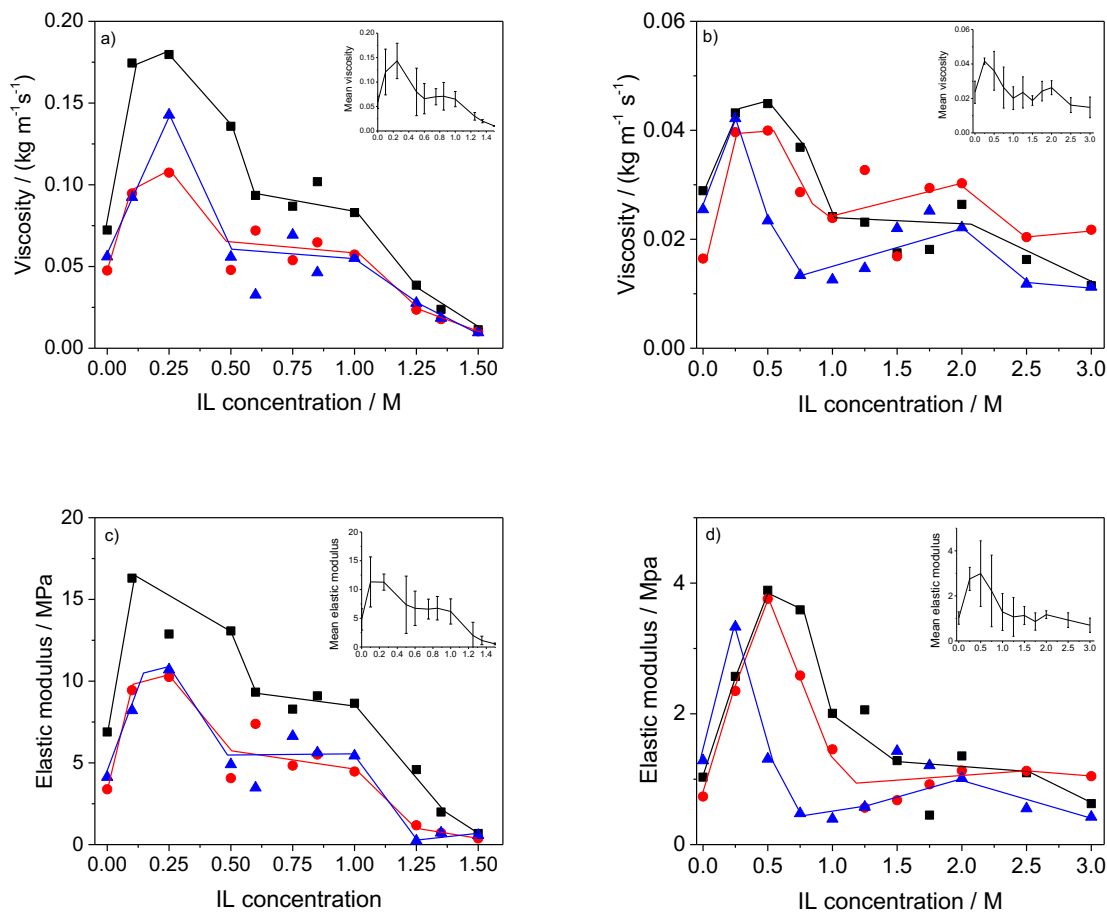

Figure 3. The viscosity of a) PEI(PSS/PDADMAC)$_4$PSS and b) PEI(PSS/PAH)$_4$PSS and the elastic modulus of c) PEI(PSS/PDADMAC)$_4$PSS and d) PEI(PSS/PAH)$_4$PSS multilayers upon swelling with IL solution of different molar concentrations. Insets show the average viscosity and elastic modulus values of the given three datasets and the corresponding standard deviations. The lines are drawn to guide the eyes.

**Discussion and Conclusions**

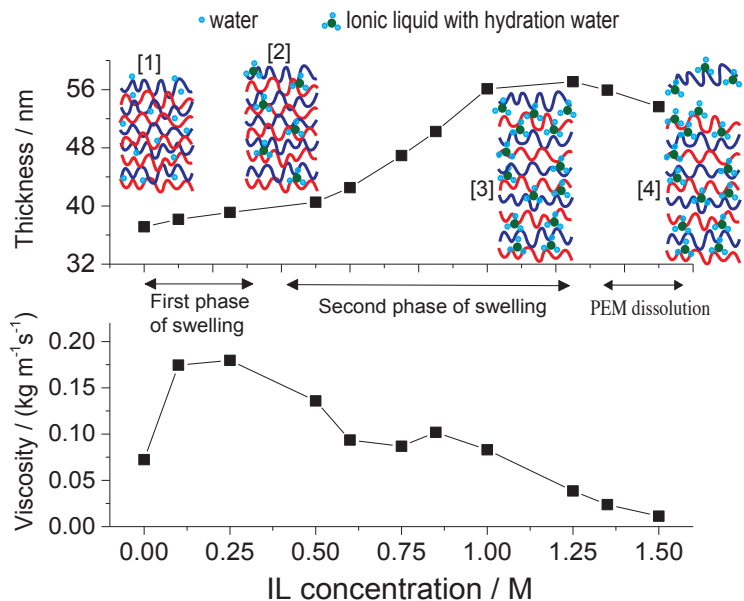

**Figure 4.** Two-phase swelling of PEI(PSS/PDADMAC)$_4$PSS with increasing ionic liquid concentration. The schematics ([1] to [4]) are drawn for visualizing the exchange of pure water to hydrated ions of IL in PEM, influencing the total water and ion content in PEM as well as thickness and viscoelasticity of PEM.

In the literature, only a handful of studies have addressed thickness, viscosity and elasticity of swollen PEM[3, 14, 22-23]. This is because of the absence of suitable experimental tools or methods to simultaneously extract these parameters from in-situ swelling experiments. The Kelvin-Voigt modeling provides a platform to extract all three parameters from a single kinetic data of QCM-D, allowing a reliable comparison of the parameters over the swelling cycles of PEM. It is to note that PEM assembled from synthetic polyelectrolytes consist roughly 40 % (v/v) water at 100 % relative humidity[4] (Scheme [1] in Figure 4). Thus, when PEM are exposed to an IL solution, a difference in the chemical potential in and out of a multilayer drives the hydrated ions of IL inside the film. The influx of IL solution reaches an equilibrium via an exchange of hydrated ions of IL and in-situ water in PEM. Because of this exchange, during the addition of IL the overall water content in PEM shifts, i.e., either increases or decreases (Figure 4). In the first phase of the swelling, the overall water content in PEM becomes lower (Scheme [2] of Figure 4). A lowered water content tends to increase the PEM density and lower the film thickness. However, the uptake of IL's ions which is relatively low at lower IL concentrations compensates this lowering in the film thickness, resulting only a minor increase in the film thickness at the first-phase of the swelling. On the other hand, a lowered hydration in PEM contributes to an increased viscosity of the multilayers and also, favorable hydrophobic interactions between the organic cation of the IL and PEM[10] tend to enhance the viscosity of the multilayer. Hence, a lower hydration and increased hydrophobic interactions upon IL uptake lead to the sharp rise in the viscoelasticity of PEM in the first-phase of the swelling (Figure 4).

At higher IL concentrations, a greater uptake of hydrated ions of IL in PEM shifts the exchange between hydrated IL and in-situ water in PEM such that the overall water content in PEM becomes higher (Scheme [3] of Figure 4). This increased hydration in PEM is one of the reasons behind the increased thickness and decreased viscoelasticity of PEM in the second phase of the swelling. It is to note that in this phase (at higher IL concentrations) a larger uptake of IL in PEM results in a greater hydrophobic interaction between IL's cation and PEM. Simultaneously, a greater electrostatic interaction between IL's ions and charge site of polyelectrolytes in PEM likely to tamper the charge-pairs of the oppositely charged polyelectrolytes, weakening the crosslinking in PEM. In other words, a higher IL uptake loosens the polymer cross-linking in PEM which also contributes to the increased PEM thickness and lowered PEM viscoelasticity. This argument is further supported by the observed irreversible swelling of PEM at higher IL concentrations (Figure S1) and partial dissolution of PSS/PDADMAC multilayer at > 1.25 M IL concentration (Figure 2 and scheme [4] of Figure 4). Thus, a balance between the hydrophobic and electrostatic interactions in the second phase of the PEM swelling influences the increasing and decreasing trend in the film thickness and viscoelasticity, respectively.

In conclusion, the article shows the application of the Kelvin-Voigt modeling in combination with measured QCM-D data to extract the thickness and viscoelasticity of swollen PEM. More importantly, the analysis of the thickness and viscoelasticity data reveals that the PEM swelling is not entirely controlled by IL uptake, rather it is influenced by the extent of the ion exchange, change in the overall water content in PEM and molecular interactions (hydrophobic and electrostatic) between PEM and IL.

ASSOCIATED CONTENT

**Supporting Information**. Figures of the measured QCM-D shifts, sensitivity of the modeling on the fluid viscosity and comparison of the modeled and measured QCM-D shifts at best-fit modeling.

AUTHOR INFORMATION

**Corresponding Author**

*Dr. Nagma Parveen; E-mail: nagma.parveen@kuleuven.be

**Funding Sources**

This work is supported by the NRW Research School "Molecules and Materials – A Common Design Principle".

**Supporting Information**

**Kelvin-Voigt model.** In the Kelvin-Voigt model a Hookean elastic spring and a Newtonian viscous dampers are connected in parallel. Equation 1 expresses the relation between the stress and the strain in a Kelvin-Voigt model. According to this model, the material deforms upon application of a constant stress and when the stress is removed the material relaxes to its original state. The elastic component stores the original state of the material whereas the viscous component causes a deformation of the material from its original state. elastic component, which stores the original state of the material and the viscous component, which causes a deformation of the material from its original state, respectively

$$\sigma_{total} = \sigma_E + \sigma_{vis} = E\varepsilon + \eta \frac{d\varepsilon}{dt} \qquad \text{Equation S1}$$

Viscoelasticity is often expressed with a complex dynamic modulus. To determine the dynamic modulus, an oscillatory stress is applied and the resulting strain is measured. The relation between the complex dynamic shear modulus and complex dynamic viscosity is given in Equation S2, where the real part of the modulus ($G'$ or $\omega\eta''$) corresponds to the elasticity or elastic modulus and the complex part of modulus ($G''$ or $\eta'$) corresponds to the viscosity. To be consistent with the literature representation, elastic modulus and viscosity are symbolized with $\mu$ and $\eta$ in the article, respectively in this article.

$$\hat{G}(\omega) = \omega G' + i\omega G'' = i\omega\hat{\eta} = i\omega\eta' + \omega\eta'' \qquad \text{Equation S2}$$

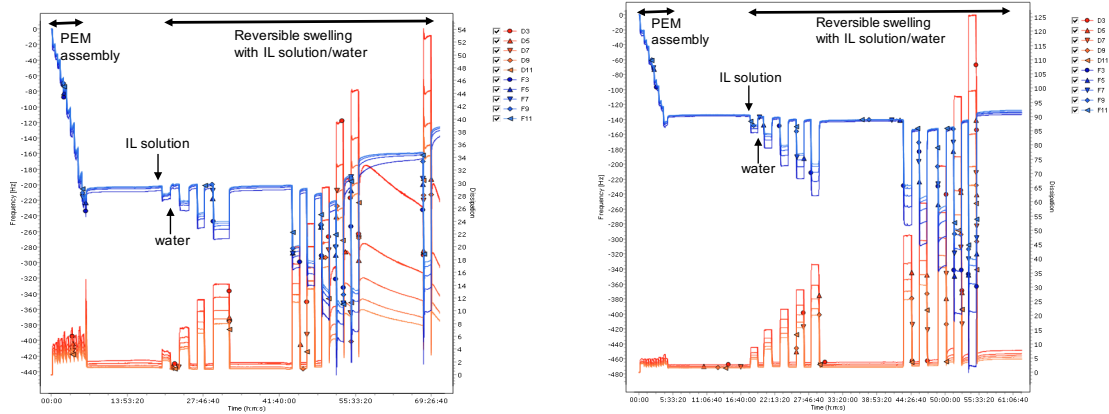

**Figure S1.** QCM-D shifts upon layer-by-layer assembly of 10 polyelectrolyte layers resulting a) PEI(PSS/PDADMAC)$_4$PSS and b) PEI(PSS/PAH)$_4$PSS multilayers. The swelling cycles were done by exposing the multilayer alternatively with ionic liquid solution and ultrapure water at successive increasing concentration of ionic liquid at 22°C. Here $D$ and $F$ stand for dissipation and frequency shifts, respectively with their corresponding overtone.

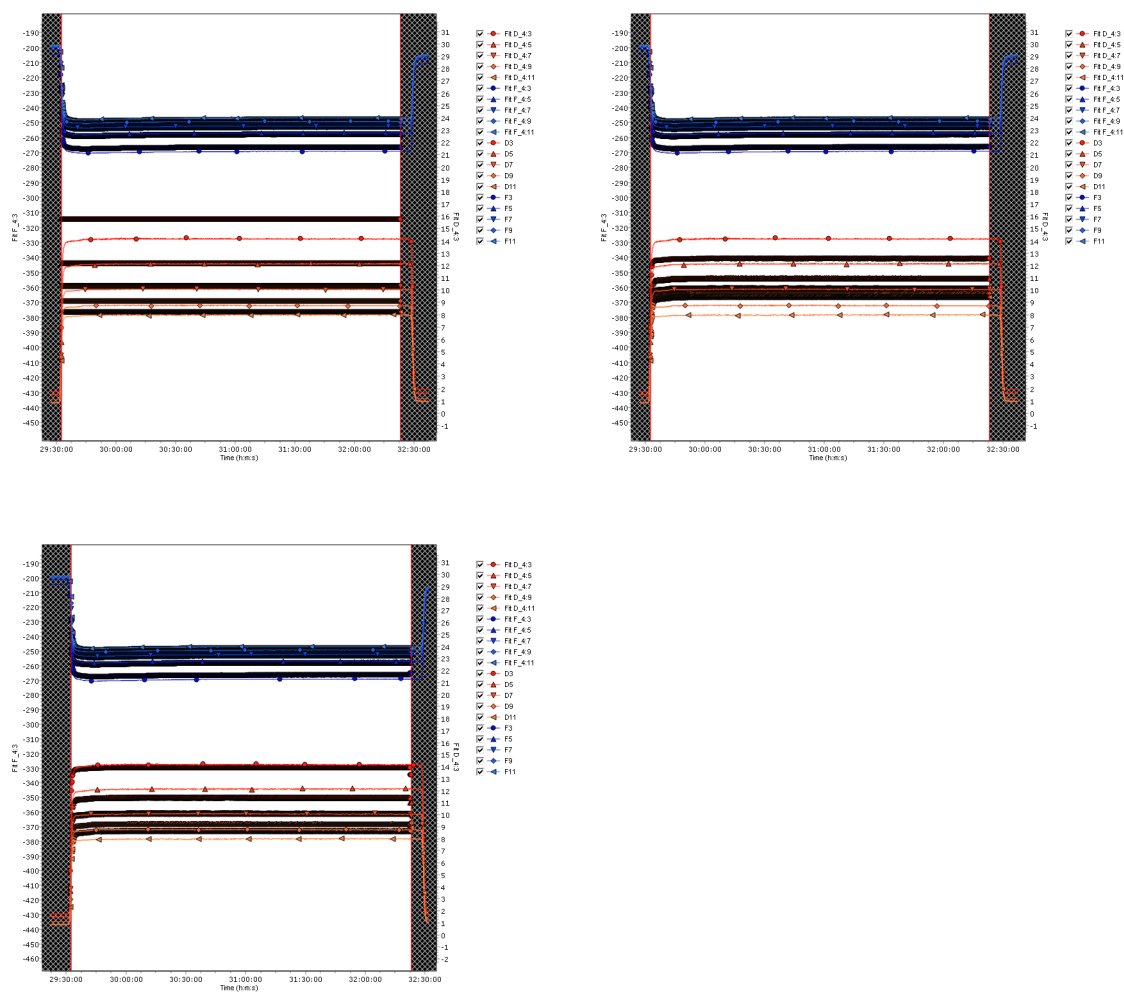

**Figure S2.** Comparison of modeled (bold dots) and measured (lines) QCM-D shifts upon swelling of PEI(PSS/PDADMAC)$_4$PSS multilayer with 0.35 M ionic liquid solution at fluid viscosity ($\eta_2$) of a) 0.0011, b) 0.0012 and c) 0.00115 kg m$^{-1}$ s$^{-1}$ without changing the values of other fixed and free parameters. These comparisons over multiple overtones clearly indicate that the best-fit is achieved at fluid viscosity of 0.00115 kg m$^{-1}$ s$^{-1}$, illustrating the sensitivity of the best-fit modeling on the viscosity of IL solution. These figures are directly taken from the Q-Tools software where the modeling was performed.

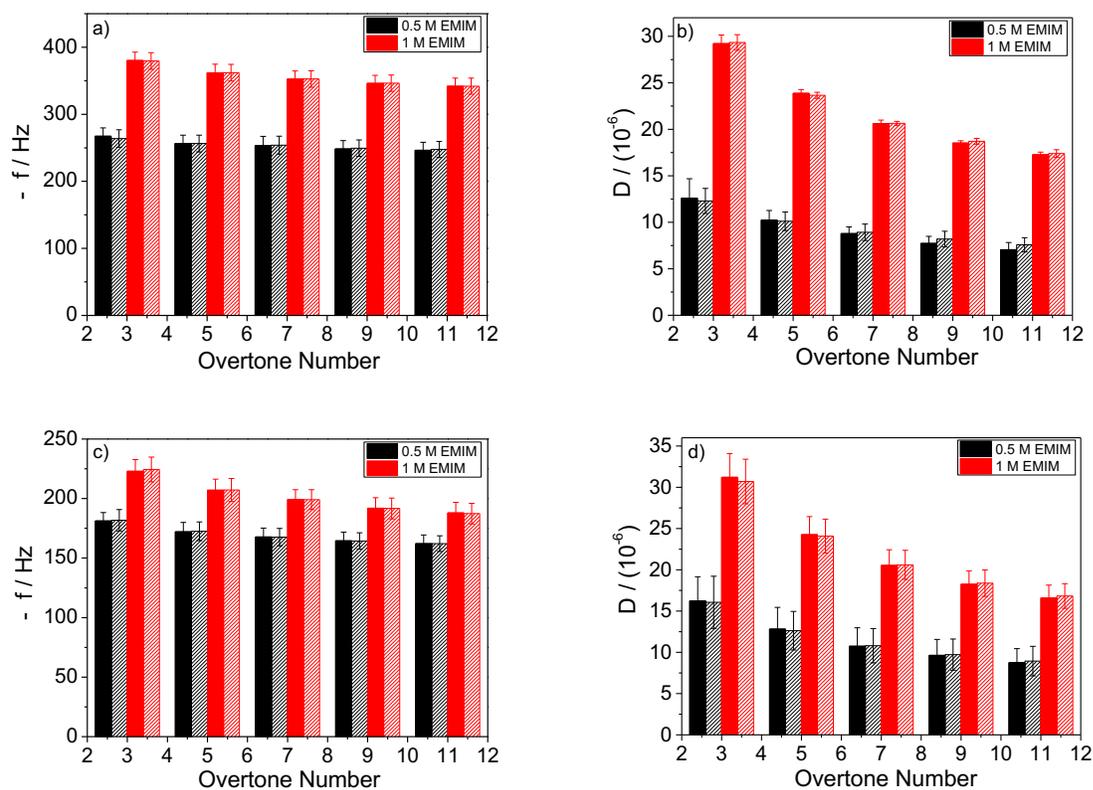

**Figure S3.** Comparison of the measured and modeled *f* and *D* shifts at best-fit modeling. The given shifts are during swelling of multilayers with 0.5 M and 1 M ionic liquid solution; a) and b) for PEI(PSS/PDADMAC)$_4$PSS and c) and d) for PEI(PSS/PAH)$_4$PSS. Dark filled and dashed filled bars represent the measured and modeled QCM-D shifts, respectively.